\documentclass[12pt,nohyper,notoc]{JHEP}

\usepackage{amsmath,amssymb,cite,graphicx} 
	
\font\zfont = cmss10 
\newcommand\ZZ{\hbox{\zfont Z\kern-.4emZ}}
\def\inbar{\vrule height1.5ex width.4pt depth0pt}
\def\IC{\relax\hbox{\kern.25em$\inbar\kern-.3em{\rm C}$}}

\newcommand{\EQ}[1]{\begin{equation} #1 \end{equation}}

\title{Quasi-Localization of Gravity by Resonant Modes}

\author{Csaba Cs\'aki$^{a,}$\footnote{J. Robert Oppenheimer Fellow.}, 
Joshua Erlich$^a$ and Timothy J. Hollowood$^{a,b}$\\
$^a$Theory Division T-8, Los Alamos National Laboratory, Los Alamos,
NM 87545, USA\\
$^b$Department of Physics, University of Wales Swansea,
Swansea, SA2 8PP, UK\\

Email: {\tt csaki@lanl.gov, erlich@lanl.gov, pyth@skye.lanl.gov}}

\abstract{We examine the behaviour of gravity in brane theories with 
extra dimensions in a non-factorizable background geometry.
We find that for metrics which are asymptotically flat far from the 
brane there is a resonant graviton mode at zero energy. The presence of this 
resonance ensures quasi-localization of gravity, whereby at intermediate 
scales the gravitational laws on the brane are approximately four dimensional. 
However, for scales larger than the lifetime of the graviton resonance 
the five dimensional laws of gravity will be reproduced due to the 
decay of the four-dimensional graviton. We also give a simple classification 
of the possible types of effective gravity theories on the brane that
can appear for general non-factorizable background geometries.
}

\preprint{{\tt hep-th/0002161}}

\begin{document}

The past two years have produced several surprising results in the field
of gravity in extra dimensions. Firstly, Arkani-Hamed, Dimopoulos and
Dvali \cite{ADD} showed that the size of compact extra dimensions could be 
as large
as a millimeter, with the fundamental Planck scale as low as a TeV,
without running into contradiction with the current short-distance gravitational
measurements, if the standard model fields are localized on a 4D ``brane''. 
Subsequently Randall and Sundrum (RS) \cite{RS} found that using a 
non-factorizable ``warped'' geometry for the extra 
dimension one could in fact have an infinitely large extra dimension,
and still reproduce Newton's Law at large distances on the brane.
The key observation of RS is that in their scenario
there is a localized zero-energy graviton bound-state in 5D which should be
interpreted as the ordinary 4D graviton. In this scenario the
geometry of the extra dimension plays a crucial r\^ole and localization
of gravity on the brane is impossible if the geometry far from the brane 
is asymptotically flat. In view of this
fact, it is even more surprising that, as Gregory, Rubakov and Sibiryakov
(GRS) \cite{GRS}
recently showed, even if the geometry is asymptotically
flat far from the brane, it is still possible to find a 
phenomenologically
viable model if one does not insist on the Newton potential
being valid at arbitrarily large scales, but instead only requires it to hold
over a range of intermediate scales (a related proposal can be found in
\cite{Kogan}). 

The aim of this paper is to present a physical explanation of the 
results of GRS and to provide a universal description of warped
gravitational theories. We will show that the reason behind the GRS
result is that even though there is no zero energy bound-state graviton
in their model, there is a resonant mode---a ``quasi bound-state''---at
zero energy, which plays the r\^ole of the 4D graviton.
The existence of this resonance at zero energy
implies the ``quasi-localization'' of gravity---that is, it tends to produce a 
region of intermediate scales on which the gravitational laws appear to
be four dimensional. We find that the long-distance scale
at which gravity appears to be five dimensional again is inversely related to the
width of the graviton resonance. The physics behind the appearance 
of this new scale is that at very large time scales the graviton decays
into plane waves away from the brane, and thus reproduces 5D 
gravity at large distances. As the width of the resonance approaches zero, the
lifetime becomes large and in the limit one regains the RS
model. On the other hand, if the resonance becomes very wide, it is 
basically washed out from the
spectrum, its effects become unimportant, and there is no longer a region where
gravity is effectively 4D.

We show that the existence of such a resonance is expected in these
kinds of theories when the geometry is asymptotically flat 
space far from the brane.  
Thus we find a simple way of classifying warped gravitational theories:
if the ground state wave function is normalizable, one has localization
of gravity \`a la Randall and Sundrum. If the ground state wavefunction
is not normalizable, but the geometry does not asymptote to flat space,
then there is simply no effective 4D gravity; however, if the ground
state wavefunction is non-normalizable and the geometry asymptotes to
flat space, we have 
quasi-localization of gravity via the resonance \`a la GRS. 

The most general 5D metric with 4D Poincar\'e symmetry can be written
\begin{equation}
ds^2\equiv g_{\mu\nu}dx^\mu\,dx^\nu=e^{-A(z)}\big(\eta_{ab}\,dx^a\,dx^b-dz^2\big)\ .
\label{confflat}\end{equation}
We will assume that the ``warp factor'' $A(z)$ is symmetric and, 
for simplicity, a non-decreasing function of $z$ for $z>0$. Furthermore we will
also assume that the matter is localized on a brane at $z=0$. 
We now consider fluctuations around the 4D Minkowski metric of the form
$h_{ab}(x,z)=e^{3A(z)/4}\psi(z)\check h_{ab}(x)$, with $-\eta^{cd}\partial_c\partial_d
\check h_{ab}(x)=m^2\check h_{ab}(x)$, where $m$ is the
four-dimensional Kaluza-Klein
mass of the fluctuation. The behaviour of the fluctuation in the
transverse space is governed by $\psi(z)$ which satisfies a Schr\"odinger-like equation:
\EQ{
-\frac{d^2\psi(z)}{dz^2}+V(z)\psi(z)=m^2\psi(z)\ ,\qquad
V(z)=\tfrac9{16}A'(z)^2-\tfrac34A''(z)\ .
\label{scheq}
}
Notice that \eqref{scheq} always admits a (not necessarily normalizable)
zero-energy wavefunction
\EQ{
\hat\psi_0(z)=\exp\big[-\tfrac34A(z)\big]\ ,
\label{zesf}
}
which potentially describes the 4D graviton.

The auxiliary quantum system described by \eqref{scheq} encodes all
the properties that we need in order to establish whether under favourable
conditions there exists an effective 4D
Newton potential on the brane. The relevant quantity to consider is the
induced gravitational potential between two unit masses on the brane. 
A discrete (normalized) eigenfunction $\psi_m(z)$ of the quantum system contributes
\EQ{
\frac{\psi_m(0)^2}{M^{3}_*}\frac{e^{-mr}}r\ ,
\label{nld}
}
to the potential, where $M_*$ is the fundamental Plank scale in 5D. 
On the other hand, continuum modes (normalized as plane waves asymptotically)
contribute 
\EQ{
\int_{m_0}^\infty dm\,\frac{\psi_m(0)^2}{M^{3}_*}
\frac{e^{-mr}}{r}\ .
\label{ccnl}
}

In the cases which we consider
the potential $V(z)$ will always approach 0 as $|z|\to\infty$ and
so $\hat\psi_0(z)$ is the only possible bound-state of the system
and the continuum begins at $m_0=0$. The potential has the
characteristic volcano shape with a central well surrounded by
barriers that decay to zero \cite{RS,RS2}.
We can distinguish 3 classes depending on the asymptotic behaviour of
$\hat\psi_0(z)$: (a) $\hat\psi_0(z)$ is
normalizable  and consequently falls off faster than $|z|^{-1/2}$; (b)
$\hat\psi_0(z)$ is non-normalizable and falls off as a power $|z|^{-\alpha}$
($\alpha\leq\tfrac12$); and (c) $\hat\psi_0(z)$ is non-normalizable
and asymptotes to a constant so the 5D spacetime is asymptotically flat.
(Note that this case requires a region where $A''(z)<0$, and therefore
cannot arise as a scalar field domain wall in an otherwise flat
background, for example.)

For case (a), $\hat\psi_0(z)$ gives rise
to the usual 4D Newton potential, as is apparent from \eqref{nld}. In
\cite{RS,US} it was shown that the effects of the continuum modes
generically  give
small corrections. The intuitive reason for this is that in order for
$\hat\psi_0(z)$ to {\em be\/} normalizable, the tunneling probability for
continuum modes of small $m$ through the barriers of $V(z)$ must
vanish as $m\to0$; this implies that $\psi_m(0)^2$ must be less singular
than $m^{-1}$. Hence the integral over the continuum modes in
\eqref{ccnl} always yields higher order corrections in $r^{-1}$
relative to the leading $r^{-1}$ piece.

For case (b) there are only continuum modes. 
In these cases $\psi_m(0)^2\sim m^{-2\alpha}$, for small $m$, and so the
potential \eqref{ccnl} behaves as $r^{2\alpha-2}$ \cite{US}. Consequently there is no
effective 4D gravity.  Note that in this case gravity does not necessarily 
appear 5D at large distances, either.

Case (c) is the main focus of interest. As in case (b) there is only
a continuum contribution to the potential \eqref{ccnl}; however, we
will argue that under favourable conditions an effective 4D Newton's
Law is recovered as in \cite{GRS} 
in some intermediate regime $r_1\ll r\ll r_2$, where
the parameters $r_{1,2}$ depend on details of the warp factor
$A(z)$. This region of ``quasi-localization'' of gravity is intimately
connected with the quasi-bound-state $\hat\psi_0(z)$ which causes a
resonance in $\psi_m(0)^2$ at $m=0$. The new long distance scale $r_2$
depends upon the width of the resonance: the narrower the resonance
the larger the scale $r_2$. In the limit in which the width goes to zero
$\hat\psi_0(z)$ becomes normalizable, $r_2\to\infty$ and 4D gravity is
obtained for all distance scales $\gg r_1$, as in the original RS
model \cite{RS}. 

\begin{figure}
\begin{center}
\includegraphics[height=6cm]{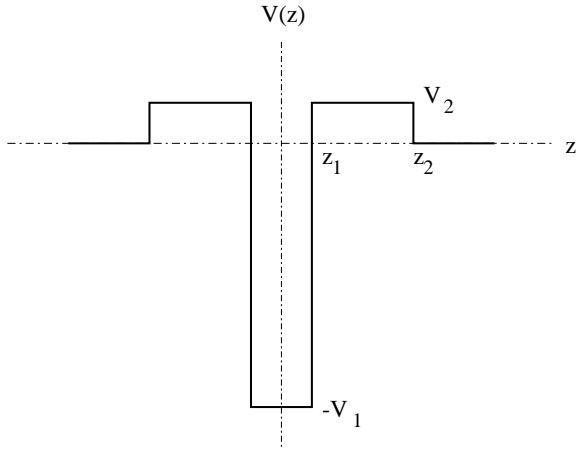}
\end{center}
\caption{\small The volcano box potential.}
\label{fig:volbox}
\end{figure}

In order to prime our intuition it is useful to consider the ``volcano
box'' potential  (Figure~\ref{fig:volbox}), for which the continuum modes can be calculated
exactly.
The solution for a symmetric continuum wavefunction 
has the form: 
\EQ{
\psi_m(z)=\frac1{\sqrt{c(m)^2+d(m)^2}}\begin{cases} \cos k_1z & |z|\leq z_1\\
a(m)\,e^{k_2(z-z_1)}+b(m)\,e^{-k_2(z-z_1)} & z_1\leq |z|\leq z_2\\
c(m)\,\cos k_3(z-z_2)+d(m)\,\sin k_3(z-z_2) & |z|\geq z_2\ ,\end{cases}
\label{psisol}
}
where 
\EQ{
k_1=\sqrt{V_1+m^2}\ ,\qquad
k_2=\sqrt{V_2-m^2}\ ,\qquad
k_3=\sqrt{m^2}\ ,
}
and
\EQ{\begin{split}
c(m)&=\cos k_1z_1\cosh k_2(z_2-z_1)-\tfrac{k_1}{k_2}\sin k_1z_1\sinh
k_2(z_2-z_1)\ ,\\
d(m)&=\tfrac{k_2}{k_3}\cos k_1z_1\sinh
k_2(z_2-z_1)-\tfrac{k_1}{k_3}\sin k_1z_1\cosh k_2(z_2-z_1)\ .
\end{split}
}
In order to have a quasi-bound-state at $m=0$ we require a solution of
the form \eqref{psisol} with $\psi_0(z)=c(0)$ for $|z|\geq z_2$. This
happens when the parameters of the potential satisfy 
\EQ{
\sqrt{V_2}\tanh\sqrt{V_2}(z_2-z_1)=\sqrt{V_1}\tan
\sqrt{V_1}z_1\ .
\label{cond}
}
The quantity we are after is
\EQ{
\psi_m(0)^2=\frac1{c(m)^2+d(m)^2}\ .
}
For small $m$, we can expand $c(m)$ and $d(m)$ in powers of $m$. The
important point is that $c(m)$ has an expansion in even powers of $m$
while $d(m)$ has a expansion in odd powers of $m$:
\EQ{
c(m)=c_0+c_1m^2+\cdots\ ,\qquad
d(m)=d_0m+d_1m^3+\cdots\ .
}
Hence for small $m$, $\psi_m(0)^2$ has the Breit-Wigner form indicative
of a resonance
\EQ{
\psi_m(0)^2=\frac{{\cal A}}{m^2+\Delta m^2}+{\cal O}(m^4)\ ,
}
where the width of the resonance is $\Delta m=|c_0|/\sqrt{d_0^2+2c_0c_1}$.
The width depends in a complicated way on the parameters of
the potential $V(z)$. A narrow resonance can be achieved by having
$c_0$ small, which requires $\sqrt{V_2}(z_2-z_1)\gg1$. In this limit
the width is approximately 
\EQ{
\Delta
m\simeq\frac{8}{(1+V_2/V_1)(z_1+1/\sqrt{V_2})}
e^{-2\sqrt{V_2}(z_2-z_1)}\ .
\label{res}
}
Intuitively, the behaviour of \eqref{res} can be deduced by the following
reasoning: in order to
get a narrow resonance we require that the tunneling
probability for modes of small $m$ through the barriers of the potential
be very small, which is achieved by having
$\sqrt{V_2}(z_2-z_1)\gg1$. In fact the exponential factor
in \eqref{res} can be deduced from a simple WKB analysis. In this
approximation the tunneling probability for the eigenfunction
$\psi_m(z)$ is
\EQ{
T(m)\thicksim\exp\Big[-2\int_{z_1}^{z_2}dz\,\sqrt{V(z)-m^2}\Big]
=\exp\Big[-2\sqrt{V_2-m^2}(z_2-z_1)\Big]
\ .
\label{wkb}
}
The width of the quasi-bound-state is $\Delta m\propto
T(0),$ giving the exponential dependence in \eqref{res}.

We expect the existence of a resonance to be generic, since it is
caused by the non-normalizable mode $\hat\psi_0(z)$. Let us suppose
that the resonance is sufficiently narrow that we can approximate
$\psi_m(0)^2$ by
\EQ{
\psi_m(0)^2=\frac{{\cal A}}{m^2+\Delta m^2}+f(m)\ ,
\label{spectrum}
}
where $f(m)$ is some underlying function which rises from 0 to 1 as
the energy goes from 0 to just over the height of the barrier. If we assume that 
$f(m)\sim m^\beta$ ($\beta>0$), for small $m$, 
the gravitational potential \eqref{ccnl} is
\EQ{
U(r)=\frac{r_2M_*^{-3}{\cal A}}{r}\int_0^\infty
dx\,\frac{e^{-xr/r_2}}{x^2+1}+{\cal O}(1/r^{\beta+2})\ ,
\label{gpot}
}
where $r_2=1/\Delta m$. The contribution from the resonance gives
the 4D Newton's Law for $r\ll r_2$ with Newton's constant
\EQ{
G_N=\frac{\pi r_2{\cal A}}{2M_*^3}\ ,
}
whereas for $r\gg r_2$ the
contribution goes as $1/r^2$ and so the 5D Newton potential is
recovered at very large distances. The rise of the 
underlying part of the continuum $f(m)$ gives the short
distance corrections to the 4D Newton's Law and sets the lower scale $r_1$.

The question now is under what conditions is the resonance narrow so that $r_2$ can be
large. To investigate this we can, following our analysis of the volcano
box potential, use the WKB approximation as a guide. The point is that
the width of the resonance is proportional to the tunneling
probability $T(m)$ for the continuum modes of zero energy, through the barrier
of the potential, evaluated at $m=0$. The WKB approximation gives
\EQ{
T(m)\thicksim\exp\Big[-2\int_{z_1}^{\infty}dz\,\sqrt{V(z)-m^2}\Big]\ ,
\label{wkbt}
}
where the integral is over the barrier region of the
potential. Notice that the integral in \eqref{wkbt} is convergent only
when $V(z)$ falls off faster than $z^{-2}$; precisely the
situation for case (c) above. In this case the limit $T(0)$ is
finite and since we expect $\Delta m\propto T(0)$ the requirement for a
narrow resonance is 
\EQ{
\int_{z_1}^{\infty}dz\,\sqrt{V(z)}\gg 1\ .
}
In other words, the barriers of the potential must be sufficiently
powerful. For case (a) above, on the other hand, the integral
diverges and $T(0)=0$, as expected since $\hat\psi_0(z)$ is normalizable.

Now we consider some concrete examples. To recap what we need in order to get
quasi-localized gravity on the brane is some geometry which is
asymptotically flat in the transverse dimension. This was realized in
the simple model of GRS \cite{GRS} by patching together 5D AdS
space onto 5D Minkowski space at some point  $z=z_0$:
\EQ{
A(z)=\begin{cases} 2\log(k|z|+1) & |z|\leq z_0\\
2\log(k|z_0|+1) & |z|\geq z_0\ .\end{cases}
}
This case can be exactly solved and one can show that quasi-localization
occurs for $k^{-1}\ll r\ll k^2z_0^3$. This allows us to check our
simple WKB analysis; in this case for $z_0\gg k^{-1}$ the WKB integral
is $\int dz
\sqrt{V(z)}\simeq\sqrt{15/4}\ln z_0$ giving $r_2\propto
z_0^{\sqrt{15}}$ (to compare with the exact behaviour $r_2\propto
z_0^3$). In the GRS scenario the patching of AdS to flat space requires the
existence of new branes at $z_0$. However, this is not a
necessary feature and one can invent scenarios which smoothly interpolate
between the AdS geometry and flat space; for instance by taking
\EQ{
A(z)=-2\log\Big(\frac1{k|z|+1}+a\Big)\ ,
\label{intg}
}
which can also be smoothed at $z=0$ by taking, for example, 
\EQ{
A(z)=-\log\Big(\frac1{k^2 z^2+1}+a^2\Big)\ .
\label{intg1}
}
In this case, the crossover occurs smoothly at $z_0\simeq 1/(ka)$.
For this example, it is a simple matter to numerically solve the
differential equation \eqref{scheq} and find $\psi_m(0)^2$ as a
function of $m$. Figure 2
illustrates this function for some values of the parameters giving
rise to a fairly broad resonance so that the rise of $\psi_m(0)^2$ to
1 can also
be seen. For small $a$ the resonance becomes very narrow and one can
verify numerically that it approximates the form in \eqref{spectrum}.
\begin{figure}
\begin{center}
\includegraphics[height=6cm]{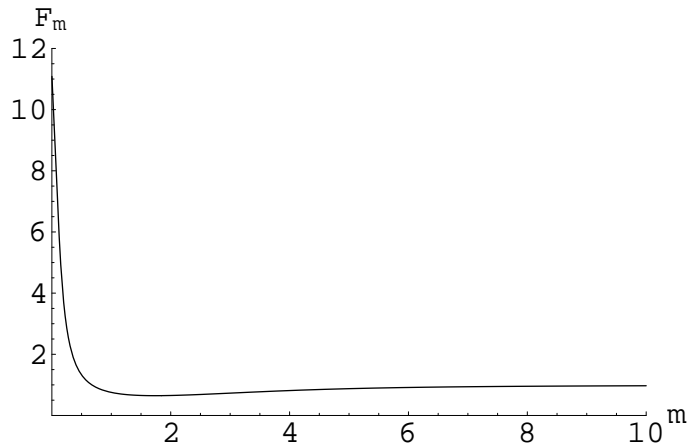}
\end{center}
\caption{\small The quantity $F_m=\frac{\pi}{2}\psi_m(0)^2$ as a 
function of $m$
for the geometry \eqref{intg1} (with $a=.25,\ k=2$), which
smoothly interpolates between AdS and Minkowski space. The resonance
at $m=0$ is clearly visible as well as the ultimate rise to 1 as the
energy goes over the height of the barrier.}
\label{fig:spect}
\end{figure}
Figure 3 shows a close-up of the resonance to illustrate the
Breit-Wigner form.
\begin{figure}
\begin{center}
\includegraphics[height=6cm]{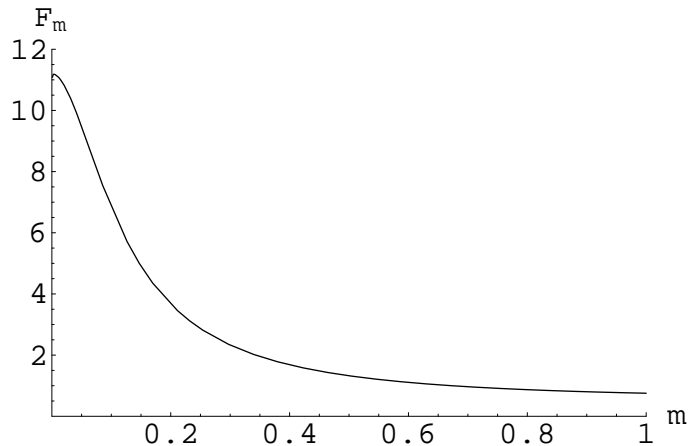}
\end{center}
\caption{\small A close-up of the resonance in Figure 2.}
\label{fig:bump }
\end{figure}
We can find the
height of the resonance from our knowledge of 
$\hat\psi_0(z)$:
\EQ{
\frac{{\cal A}}{\Delta
m^2}=\frac{2}{\pi}\hat\psi_0(0)^2=e^{-3(A(0)-A(\infty))/2}\simeq 
\frac{2}{\pi}a^{-3}\ .
\label{relo}
}
In addition, when the resonance is very narrow, we can find the
width to leading order in $a$ 
by using the fact that,
as $a\to0$, $\hat\psi_0(z)$ becomes normalizable and the effect of the
resonance should appproximate a delta function. This gives
\EQ{
\frac{{\cal A}}{\Delta m}=\frac{4k}{\pi}\ .
\label{relt}
}
Hence, $\Delta m\simeq 2ka^3\simeq 2 k^2z_0^{-3}=r_2^{-1}$. 

We have introduced the idea of ``quasi localization'' of gravity on
the brane as the general notion lying behind the scenario of GRS \cite{GRS} 
whereby
over a large range of intermediate distance scales the gravitational potential
is, to a good accuracy, Newtonian. We argued that the relevant geometry
for quasi localization is when the transverse geometry becomes
asymptotically flat. The new large scale above which the
gravitational potential becomes five-dimensional depends on the scale
at which the crossover to flat space occurs. Physically we can
describe the onset of 5D gravity on the brane at this new scale by
saying that the effective 4D graviton is unstable and decays into the KK
continuum. Obviously in order to be phenomenologically viable the
lifetime must be very long. Although we have only considered the case
with one extra dimension we expect that the same phenomenon to occur
with any number of extra dimensions.
In the same vein, we note that quasi-localization may be of relevance in string
theory where the transverse geometry to a 
$p$-brane soliton is indeed asymptotically flat.

We would like to thank Yuri Shirman for useful discussions and fruitful
collaboration.
C.C. is an Oppenheimer Fellow at the 
Los Alamos National Laboratory.
C.C., J.E. and T.J.H. are supported 
by the US Department of Energy under contract W-7405-ENG-36.

{\it Note added:} After this paper has been completed a revised version
of Ref.~\cite{GRS} (as well as Ref.~\cite{Dvali}) appeared, which also 
independently noted the presence of the resonance in these theories.

\end{document}